\documentclass[runningheads]{llncs}

\usepackage{multirow}
\usepackage{graphicx}
\usepackage[table,xcdraw]{xcolor}
\usepackage{booktabs}
\usepackage{amsmath}
\usepackage{tikz}
\usepackage{rotating}
\usepackage{float}
\usepackage{amssymb}

\usepackage[normalem]{ulem}
\useunder{\uline}{\ul}{}
\newcommand{\ftextbf}[1]{%
    \pdfliteral direct {2 Tr 0.3 w} 
     #1%
    \pdfliteral direct {0 Tr 0 w}%
}

\begin{document}
\definecolor{azzurrino}{RGB}{218, 227, 243}

\title{Review of data types and model dimensionality for cardiac DTI SMS-related artefact removal\thanks{This work was supported in part by the UKRI CDT in AI for Healthcare \url{http://ai4health.io} (Grant No. EP/S023283/1), the British Heart Foundation (RG/19/1/34160), and the UKRI Future Leaders Fellowship (MR/V023799/1).}}

\titlerunning{Review of data types and model dimensionality for cardiac DTI}


\author{Michael T\"{a}nzer\inst{1}\orcidID{0000-0002-9046-1008} \and
Sea Hee Yook\inst{1,2} \and
\\Pedro Ferreira\inst{1,2}\orcidID{0000-0002-0436-3496} \and
\\ Guang Yang*\inst{1,2}\orcidID{0000-0001-7344-7733} \and
\\ Daniel Rueckert*\inst{1,3}\orcidID{0000-0002-5683-5889} \and
\\ Sonia Nielles-Vallespin*\inst{1,2}\orcidID{0000-0003-0412-2796}}

\authorrunning{M. T\"{a}nzer et al.}

\institute{Imperial College London \and
Royal Brompton and Harefield hospital \and
Klinikum rechts der Isar, Technische Universität München (TUM) \\ Send correspondence to \{m.tanzer, g.yang\}@imperial.ac.uk \\ *Co-last senior authors}

\maketitle

\begin{abstract}
    As diffusion tensor imaging (DTI) gains popularity in cardiac imaging due to its unique ability to non-invasively assess the cardiac microstructure, deep learning-based Artificial Intelligence is becoming a crucial tool in mitigating some of its drawbacks, such as the long scan times. As it often happens in fast-paced research environments, a lot of emphasis has been put on showing the capability of deep learning while often not enough time has been spent investigating what input and architectural properties would benefit cardiac DTI acceleration the most. In this work, we compare the effect of several input types (magnitude images vs complex images), multiple dimensionalities (2D vs 3D operations), and multiple input types (single slice vs multi-slice) on the performance of a model trained to remove artefacts caused by a simultaneous multi-slice (SMS) acquisition. Despite our initial intuition, our experiments show that, for a fixed number of parameters, simpler 2D real-valued models outperform their more advanced 3D or complex counterparts. The best performance is although obtained by a real-valued model trained using both the magnitude and phase components of the acquired data. We believe this behaviour to be due to real-valued models making better use of the lower number of parameters, and to 3D models not being able to exploit the spatial information because of the low SMS acceleration factor used in our experiments.
\end{abstract}

\section{Introduction}
    Cardiac diffusion tensor imaging is growing as a novel imaging modality as it is capable of interrogating the microstructure of the beating heart with no invasive surgery and without the use of any contrast agent \cite{mori_principles_2006}, making it accessible for patients with reduced kidney functionality \cite{schlaudecker_gadolinium-associated_2009} or for frequent scans. In clinical research studies, cardiac DTI has been shown to be useful in phenotyping several cardiomyopathies such as hypertrophic cardiomyopathy (HCM) and dilated cardiomyopathy (DCM) by quantitatively analysing the microstructural organisation and orientation of cardiomyocytes within the myocardium.
    
    As cardiac DTI is becoming more and more studied, the use of deep learning-based (DL) approaches applied to it is similarly increasing \cite{phipps_accelerated_2021,schlemper_stochastic_2018,ferreira_automating_2020,ferreira_accelerating_nodate,karimi_diffusion_2022,weine_synthetically_2022,cao_cs-gan_2020,tanzer_faster_2022}. Most of the recent work in the field, although, suffer from a common crucial shortcoming: in order to quickly show the potential of deep learning on improving cardiac DTI, many publications don't go beyond applying some well-known general-purpose architecture to the data they have available, often ignoring inherent properties of the acquisition method. As an example, in much of the recently published work, we find a widespread use of out-of-the-box models such as the popular U-Net \cite{ronneberger_u-net_2015}, which is a 2D model designed for real data. MRI data, on the other hand, is complex by definition and the subject of the acquisition is often 3D rather than 2D, making the standard U-Net ill-fitting for the task.
    
    In this work, we compare the effect of making relatively small architectural changes to the popular U-Net model when applied to a general image-to-image task in DTI. Specifically, we choose the task of removing artefacts from cardiac DTI images acquired with an SMS protocol, and we compare the classic 2D magnitude-only U-Net with 3D and complex versions of the same model, with the goal of providing future researchers with a better starting point for their experimental work. To this end, we also make the code for all of our tested models available on GitHub\footnote{\url{https://github.com/Michael-Tanzer/architectures-tanzer-stacom22}}.

\section{Background}
    \subsection{Cardiac diffusion tensor imaging}
    Diffusion tensor imaging measures the diffusion of water molecules for every voxel in the imaged tissue and approximates as a 3D tensor. As the free diffusion of water in the tissue is constrained by the shape of cardiac muscle microstructure, studying such tensors has been shown to give us information related to the shape and orientation of the cardiomyocytes in the imaged tissues.
    
    The cardiac diffusion tensor information is commonly visualised and quantified through four per-voxel metric maps: Mean Diffusivity (MD) that quantifies the total diffusion in the voxel (higher corresponds to more diffusion), Fractional Anisotropy (FA) that quantifies the level of organisation of the tissue (higher corresponds to a higher organisation), Helix Angle (HA) and Second Eigenvector (E2) Angle (E2A) that quantify the 3D orientation and shape of the tissue in the voxel \cite{basser1995inferring,kung2011presence}. These maps have been shown to be a promising tool for phenotyping many cardiac pathology in a clinical setting \cite{bihan_diffusion_2001,niellesvallespin_cardiac_2020,nielles-vallespin_assessment_2017}.
    
    \subsection{Simultaneous multi-slice acquisition (SMS)}
    Simultaneous multi-slice techniques have been used with great success to reduce the acquisition time in brain diffusion tensor imaging (DTI) \cite{setsompop_improving_2012}. SMS uses a multi-band excitation pulse to simultaneously excite several 2D slices within the imaged tissue. In SMS, each receiver coil collects a single frequency-domain image for all the excited slices, the signal received is a weighted sum of the signals that would be emitted by exciting each slice individually.
    
    By making use of the redundant information from the receiver coils that surround the tissue we want to image, we can then separate the signal from the excited slices using modified versions of the GRAPPA \cite{griswold_generalized_2002} and SENSE \cite{pruessmann_sense_1999} algorithms used for in-plane acceleration. Unfortunately, as the information is often insufficient and as the problem is not fully characterised, the slice separation algorithms introduce artefacts in the separated slices \cite{barth_simultaneous_2016}. The artefacts are often referred to as interslice-leakage artefact as they arise when information from one slice erroneously ends up on a different slice. Because of the disposition in space of the MRI coils, the leakage between two slices is more evident when the slices are closer in the imaged tissue.
    
    \subsection{Deep learning in DTI}
    As cardiac DTI grows in popularity, a lot of work has been published trying to improve the acquisition quality or trying to shorten its long scan-times. Among others, Ferreira et al. \cite{ferreira_accelerating_nodate}, T\"{a}nzer et al. \cite{tanzer_faster_2022}, and Phipps et al. \cite{phipps_accelerated_2021} reduced the number of repetitions needed to increase the low SNR by using a de-noising framework to restore high image quality from less acquired data. Schlemper et al. \cite{schlemper_stochastic_2018} applied a cascade of convolutional neural networks to fill-in k-space entries acquired with a compressed sensing protocol, further reducing the scan times.
    
    Ferreira et al. \cite{ferreira_automating_2020} propose a U-Net for the segmentation of the left ventricle, automating part of the DTI maps computation, Cao et al. \cite{cao_cs-gan_2020} show how a GAN model can be used as a de-aliasing model for DTI, and Tian et al. \cite{tian_deep_2020} work on self-supervised DTI de-noising is also based on a modified U-Net model.
    
    Most of the examples reported above, despite all being a great contribution to the field of deep-learning-accelerated cardiac DTI, have a 2D real-valued U-Net model at their core. This shows that the choice of data type and dimensionality is not the top priority of many influential publications.
    

\section{Methods}
    \subsection{Complex neural networks}
    As MRI data is inherently complex, we explore the possibility of training using complex data. Traditionally, there are two main ways to achieve this: separating the complex data into non-complex components, or using a model that performs complex operations.
    
    The former is a more straight-forward approach: the complex data is split into its real and imaginary components or into magnitude and phase and then a real-values model is trained using the split data. This can be done in multiple ways: either by treating the components as different channels, or by training two separate models, one for each component. In our comparison we split the data in magnitude and phase as they are more meaningful in the physics of MRI and we use them as separate input channels to a unified model.
    
    The latter option requires more thinking: while some operations naturally extend to the complex domain, some others do not and need to be re-designed. We report the main changes to the used operators in Table \ref{tab:complex-nns}. In the table $z = a + ib$ where $i = \sqrt{-1}$, \textit{c} is the convolution layer, and \textit{tc} is the transpose convolution layer.
    
    	\begin{table}[tbh]
		\centering
		\footnotesize
		\begin{tabular}{lllll}
			\toprule
			\multicolumn{2}{c}{Operation}                         & Naive imp.                 & Complex equivalent                                                  \\ \midrule \midrule
			Multiplication    & Inner product $\ $ & $\sum_i w_i z_i$             & \checkmark                                                                     \\
			& Convolution            & $\operatorname{c}(a+ib)$                   & $ \operatorname{c}_r (a) -  \operatorname{c}_i(b) + i ( \operatorname{c}_i (a) +  \operatorname{c}_r (b))$    \\
			& Trans. conv.  & $ \operatorname{tc}(a+ib)$                 &  $ \operatorname{tc}_r (a) + \operatorname{tc}_i(b) + i ( \operatorname{tc}_i (a) -  \operatorname{tc}_r (b))$ \\ \midrule
			Activation        & Sigmoid                & $\frac{1}{1+e^{-(a + ib)}}$  &  $\operatorname{sigmoid}(a) + i\operatorname{sigmoid}(b)$                             \\
			&
			ReLU &
			$\max(0, a + ib)$ &
			\begin{tabular}[c]{@{}l@{}}$\left\{\begin{array}{ll} (|z|+b) \frac{z}{|z|} & \text { if }|z|+b \geq 0 \\ 0 & \text { if }|z|+b<0\\ \end{array}\right.$\end{tabular} \\
			& Dropout                & $\operatorname{DO}(a) + i \operatorname{DO}(b)\ \ $ & $\operatorname{DO}(a + ib)$                                      \\ \midrule
			Pooling           & Max                    & $\max_i z_i$                 & $z_k$ where $\operatorname{argmax}_k\{|z_k|\}$                          \\
			& Average                & $\frac{1}{N} \sum_i z_i$    & \checkmark                                                              \\ \midrule
			Normalisation $\ $ &
			Batch norm &
			$\mathbf{\hat{z}} = \frac{\mathbf{z} - \mathbb{E}[\mathbf{z}]}{\sqrt{\mathbb{V}[\mathbf{z}] + \epsilon}}$ &
			\checkmark\\ \midrule
			Loss              & Euclidean              & $|| Y - Z ||^2_2$              & $|| \operatorname{abs}(Y) - \operatorname{abs}(Z) ||_2^2$                                                                   \\
			& L1                     & $|| Y - Z ||_1$           & $|| \operatorname{abs}(Y) - \operatorname{abs}(Z) ||_1$                                                                   \\ \bottomrule\\
			\end{tabular}
			\caption{Neural networks operators and their complex counterparts. Marked with a checkmark those operations that don't need substantial modification to extend their usage to complex space.}
			\label{tab:complex-nns}
		\end{table}
		
	Once we have defined these basic operations we can then build a model that uses them and is therefore fit to process complex numbers. The advantage of this approach is that the model makes use of properties of complex numbers during the training instead of letting the network learn a relationship between the input channels like in the previous method.
    
    \subsection{Experimental setting}
    In order to provide a better understanding on the effect of using a 3D or complex model, we compare all combinations of architectures obtained by modifying the following properties:
    \begin{itemize}
        \item 2D vs 3D: whether the model uses 2D operations or 3D operations.
        \item Magnitude vs complex data: whether we train the model with the absolute component alone or whether we use the full complex representation of the data. Notice that ``complex data" is further split into 1. fully complex models that use complex operations and 2. models that keep standard operations, but use the phase data as a separate input channel. We refer to the former magnitude-only models as ``Mag", fully complex models as ``Comp", and magnitude-and-phase models as ``MagPhs" for brevity.
        \item All slices vs individual slices: whether the model is trained to correct all SMS-acquired slices simultaneously or whether to correct a single slice at a time. When all the slices are used and the model is 3D, the slices are arranged in the third spatial dimension, while for 2D models the slices are treated as image channels.
    \end{itemize}
    This comparison results in 12 combinations of dimensionality (2 types), data (3 types), and input data (2 types). When a single model is referred to, we often use a shorthand: for instance, if we refer to a 2D fully-complex model trained on all the SMS slices, we will shorten it as ``2D-All-Comp".

    All the models were trained for 200 epochs with the Adam optimiser \cite{kingma_adam_2017}, a learning rate of 0.0003 that was then lowered by a factor of 10 after 100 epochs, a batch size of 16, mean absolute error loss, and residual learning. The data was padded and normalised in the range 0 to 1 and then randomly augmented with random rotation and random vertical and horizontal flipping. All the results were computed on the test set for the epoch in which the validation MAE was lowest.
    
    In order to ensure consistency, we kept the model size and architecture as fixed as possible by choosing a number of parameters, 3 millions, and a general architecture (U-Net with 5 layers with a doubling number of channels in each encoding layer) and subsequently adjusting the initial number of channels. The models have the following starting number of channels:
    \begin{itemize}
        \item 2D Mag and MagPhs: 28
        \item 3D Mag and MagPhs: 16
        \item 2D Complex: 20
        \item 3D Complex: 11
    \end{itemize}
    
    The data used for the training came from 31 ex-vivo swine hearts and was acquired with SMS factor equal to 2 and distance factor equal to 400\%. Each heart was scanned in multiple locations to cover as much of the volume as possible. The ground truth images were obtained by scanning each heart again with the same protocol but no SMS acceleration. This results in around 43,000 2D complex slices for the training, 1200 slices for validation and 1200 slices for testing.
    
    \subsection{Results evaluation}
    When working with DTI data, there are two main components we are interested in evaluating: the acquired images and the DTI maps derived from the images. The latter are particularly important as they are the main tools a clinician would use in a clinical setting. When evaluating the artefact-removal results of our models we therefore need to take both into account. 
    
    \noindent To evaluate the image quality we use mostly standard wide-spread metrics:
    \begin{itemize}
        \item Mean Absolute Error (MAE) ↓: $\frac{1}{nm}\sum_{i=1}^n \sum_{j=1}^m \left| X^{(i, j)} - Y^{(i, j)} \right|$ where $x$ and $y$ are the predicted and target images respectively. For complex images the MAE is computed with respect to their magnitude and for MagPhs the phase information is not taken into account.
        \item Peak Signal to Noise Ratio (PSNR) ↑: $10 \log_{10}\left(\frac{\operatorname{MAX}_X^2}{\operatorname{MSE}}\right)$ where $\operatorname{MAX}_X$ is maximum pixel intensity value across the image and $\operatorname{MSE}$ is the mean squared error between the predicted image and the target image. In the case of complex images we compute the PSNR on their magnitude and for the MagPhs case we discard the phase data.
        \item Structural Similarity Index (SSIM) ↑: SSIM measures the perceived image degradation based on the loss of structure between the output and target images. It is computed as follows:
        \begin{equation}
            \operatorname{SSIM}(X, Y)=\frac{\left(2 \mu_{X} \mu_{X}+c_{1}\right)\left(2 \sigma_{XY}+c_{2}\right)}{\left(\mu_{X}^{2}+\mu_{Y}^{2}+c_{1}\right)\left(\sigma_{X}^{2}+\sigma_{X}^{2}+c_{2}\right)}
        \end{equation}
        Where $\mu_I$ represents the mean over $I$, $\sigma_I$ the standard deviation over $I$, $\sigma_{IJ}$ the covariance, and $c_1$ and $c_2$ are fixed scalars used for numerical reasons. 
    \end{itemize}
    
    When analysing the DTI maps, we need to distinguish between scalar maps (MD and FA) and angular maps (HA and E2A). While for scalar maps we can use MAE (↓) as an error metric, angular maps are defined in the range $[-90^\circ, 90^\circ)$ and they wrap around at the two extrema of this range. For these maps we use the Mean Angle Absolute Error (MAAE, ↓) defined below instead
    \begin{equation*}
        \operatorname{MAAE}(X, Y) = \frac{1}{NM} \sum_{i=0}^{N} \sum_{j=0}^{M} \begin{cases}
        \left|X^{(i,j)} - Y^{(i,j)}\right|,\ if\ \left|X^{(i,j)} - Y^{(i,j)}\right| < 90^\circ\\
        180^\circ - \left|X^{(i,j)} - Y^{(i,j)}\right|,\ \text{otherwise}
        \end{cases}
    \end{equation*}
    Moreover, as the DTI maps are not well-defined for the background voxels, we only consider the metrics computed on voxels belonging to the cardiac tissue.
    
    All the results are reported as the median over the means on a per-slice basis and its inter-quartile range as \textit{median [iqr]}. When we perform a statistical significance test we use the Wilcoxon rank test with $P=0.05$.

\section{Results}
    In Tables \ref{tab:numerical-results-maps} and \ref{tab:numerical-results-images} we report the performance metrics for the output images and for the DTI maps on the test set for all our models. The values of MD have been scaled by $10^5$ and the values of FA by $10^2$ to improve readability. In the table we mark in bold the best result across all models and we underline the second-best result. The values are also colour-coded from green (best result) to red (worst result) through yellow (median result) on a per-metric basis to better compare the models at first glance.
    
    \begin{table}[tbh]
    \footnotesize
    \centering
    \begin{tabular}{@{}l|rrrrrrrr@{}}
    \toprule
                               & \multicolumn{2}{c}{HA}                            & \multicolumn{2}{c}{E2A}                     & \multicolumn{2}{c}{MD ($\times10^5$)}               & \multicolumn{2}{c}{FA ($\times10^2$)}                     \\ \cmidrule(l){2-9} 
    \multirow{-2}{*}{Run name} & Median                                      & IQR & Median                                & IQR & Median                                & IQR  & Median                                      & IQR  \\ \midrule
    2D-ALL-ABS                 & \cellcolor[HTML]{E0E5B6}16.9                & 5.0 & \cellcolor[HTML]{E8E9BC}26.1          & 4.8 & \cellcolor[HTML]{F4EDC4}5.44          & 1.51 & \cellcolor[HTML]{FFCDB1}5.95                & 1.19 \\
    2D-ALL-Comp                & \cellcolor[HTML]{FFE8C5}18.0                & 4.6 & \cellcolor[HTML]{F2EDC3}26.4          & 4.7 & \cellcolor[HTML]{FBF0C9}5.50          & 0.97 & \cellcolor[HTML]{A9D08E}\ftextbf{5.41}      & 1.33 \\
    2D-ALL-MagPhs              & \cellcolor[HTML]{DCE4B2}16.7                & 2.6 & \cellcolor[HTML]{A9D08E}\ftextbf{24.5}& 4.4 & \cellcolor[HTML]{FFEFCA}5.55          & 1.03 & \cellcolor[HTML]{FFEDC9}5.86                & 0.40 \\
    3D-ALL-ABS                 & \cellcolor[HTML]{E1E6B6}16.9                & 4.3 & \cellcolor[HTML]{F4EDC4}26.4          & 3.5 & \cellcolor[HTML]{FFAA98}5.98          & 1.26 & \cellcolor[HTML]{FF6565}6.24                & 0.59 \\
    3D-ALL-Comp                & \cellcolor[HTML]{FF6565}19.4                & 4.8 & \cellcolor[HTML]{FF6E6B}27.8          & 5.1 & \cellcolor[HTML]{FF897F}6.19          & 1.47 & \cellcolor[HTML]{FFB09C}6.03                & 3.26 \\
    3D-ALL-MagPhs              & \cellcolor[HTML]{A9D08E}\ftextbf{15.0}      & 3.8 & \cellcolor[HTML]{D1DFAA}{\ul 25.5}    & 4.8 & \cellcolor[HTML]{B1D394}{\ul 4.89}    & 0.67 & \cellcolor[HTML]{B2D395}{\ul 5.46}          & 1.35 \\
    2D-Single-ABS              & \cellcolor[HTML]{D3E0AC}{\ul 16.4}          & 4.2 & \cellcolor[HTML]{D1DFAA}{\ul 25.5}    & 5.5 & \cellcolor[HTML]{DEE5B4}5.26          & 1.28 & \cellcolor[HTML]{FCF0C9}5.83                & 0.51 \\
    2D-Single-Comp             & \cellcolor[HTML]{FFAF9B}18.6                & 4.7 & \cellcolor[HTML]{FFAE9A}27.3          & 5.0 & \cellcolor[HTML]{D9E3B1}5.22          & 0.95 & \cellcolor[HTML]{D4E1AD}5.63                & 2.25 \\
    2D-Single-MagPhs$\ \ $     & \cellcolor[HTML]{FF9487}18.9                & 3.8 & \cellcolor[HTML]{FFB59F}27.2          & 4.7 & \cellcolor[HTML]{FFD5B7}5.71          & 1.17 & \cellcolor[HTML]{FFD4B6}5.93                & 0.37 \\
    3D-Single-ABS              & \cellcolor[HTML]{FBF0C9}17.8                & 3.1 & \cellcolor[HTML]{FFD2B5}27.0          & 5.4 & \cellcolor[HTML]{FF6565}6.41          & 1.92 & \cellcolor[HTML]{EEEBBF}5.76                & 0.34 \\
    3D-Single-Comp             & \cellcolor[HTML]{FF9487}18.9                & 5.2 & \cellcolor[HTML]{FFCBB0}27.0          & 5.3 & \cellcolor[HTML]{A9D08E}\ftextbf{4.82}& 0.96 & \cellcolor[HTML]{B6D597}5.48                & 1.12 \\
    3D-Single-MagPhs           & \cellcolor[HTML]{FFC8AD}18.4                & 3.2 & \cellcolor[HTML]{FF6565}27.9          & 6.0 & \cellcolor[HTML]{FF7A75}6.28          & 1.88 & \cellcolor[HTML]{FFE6C3}5.88                & 0.72 \\ \bottomrule
    \multicolumn{5}{c}{}
    \end{tabular}
    \caption{Numerical results related to the DTI maps computed from the AI-processed data. We report MAE for MD and FA and MAAE for HA and E2A. In bold and underlined, respectively, the best and second-best results for each metric.}
    \label{tab:numerical-results-maps}
    \end{table}

    \begin{table}[tbh]
    \footnotesize
    \centering
    \begin{tabular}{@{}l|rrrrrr@{}}
    \toprule
                               & \multicolumn{2}{c}{MAE ($\times10^3$)}                    & \multicolumn{2}{c}{PSNR}                     & \multicolumn{2}{c}{SSIM}                       \\ \cmidrule(l){2-7} 
    \multirow{-2}{*}{Run name} & Median                                      & IQR  & Median                                & IQR  & Median                                 & IQR   \\ \midrule
    2D-ALL-ABS                 & \cellcolor[HTML]{FFF2CC}1.75                & 0.73 & \cellcolor[HTML]{BED89D}{\ul 37.2}    & 3.56 & \cellcolor[HTML]{FFF1CB}0.917          & 0.021 \\
    2D-ALL-Comp                & \cellcolor[HTML]{FFF1CB}1.82                & 0.78 & \cellcolor[HTML]{FFF1CB}36.6          & 3.26 & \cellcolor[HTML]{FFEEC9}0.911          & 0.026 \\
    2D-ALL-MagPhs              & \cellcolor[HTML]{FFF0CB}1.84                & 0.78 & \cellcolor[HTML]{A9D08E}\ftextbf{37.4}& 2.85 & \cellcolor[HTML]{DFE6B5}0.921          & 0.020 \\
    3D-ALL-ABS                 & \cellcolor[HTML]{F9EFC7}1.74                & 0.62 & \cellcolor[HTML]{DBE4B2}37.0          & 2.80 & \cellcolor[HTML]{EDEBBF}0.920          & 0.018 \\
    3D-ALL-Comp                & \cellcolor[HTML]{FF6565}7.15                & 3.92 & \cellcolor[HTML]{FF9E8F}31.1          & 1.41 & \cellcolor[HTML]{FF6565}0.618          & 0.012 \\
    3D-ALL-MagPhs              & \cellcolor[HTML]{DAE3B1}1.71                & 0.71 & \cellcolor[HTML]{FFD8B9}34.9          & 3.94 & \cellcolor[HTML]{CCDEA8}0.922          & 0.020 \\
    2D-Single-ABS              & \cellcolor[HTML]{BDD89C}1.68                & 0.76 & \cellcolor[HTML]{C0D99F}37.2          & 4.52 & \cellcolor[HTML]{BFD99E}{\ul 0.923}    & 0.022 \\
    2D-Single-Comp             & \cellcolor[HTML]{FFEBC7}2.04                & 0.75 & \cellcolor[HTML]{FFECC7}36.2          & 3.26 & \cellcolor[HTML]{FFE5C3}0.892          & 0.024 \\
    2D-Single-MagPhs$\ \ $     & \cellcolor[HTML]{AED291}{\ul 1.67}          & 0.75 & \cellcolor[HTML]{FDF2CB}36.6          & 5.31 & \cellcolor[HTML]{A9D08E}\ftextbf{0.925}& 0.022 \\
    3D-Single-ABS              & \cellcolor[HTML]{A9D08E}\ftextbf{1.66}      & 0.72 & \cellcolor[HTML]{FF6565}27.2          & 5.46 & \cellcolor[HTML]{FFEDC8}0.909          & 0.028 \\
    3D-Single-Comp             & \cellcolor[HTML]{FF978A}5.24                & 4.19 & \cellcolor[HTML]{FFEBC7}36.2          & 3.37 & \cellcolor[HTML]{FFE4C2}0.889          & 0.020 \\
    3D-Single-MagPhs           & \cellcolor[HTML]{D3E0AC}1.70                & 0.77 & \cellcolor[HTML]{CCDEA7}37.1          & 3.83 & \cellcolor[HTML]{C0D99F}{\ul 0.923}    & 0.026 \\ \bottomrule
    \multicolumn{5}{c}{}
    \end{tabular}
    \caption{Numerical results related to the artefact-removal output of our proposed AI models. The results here refer to the images produced by our model.  In bold and underlined, respectively, the best and second-best results for each metric.}
    \label{tab:numerical-results-images}
    \end{table}

\begin{figure}[tbh]
        \centering
        \includegraphics[width=0.85\textwidth]{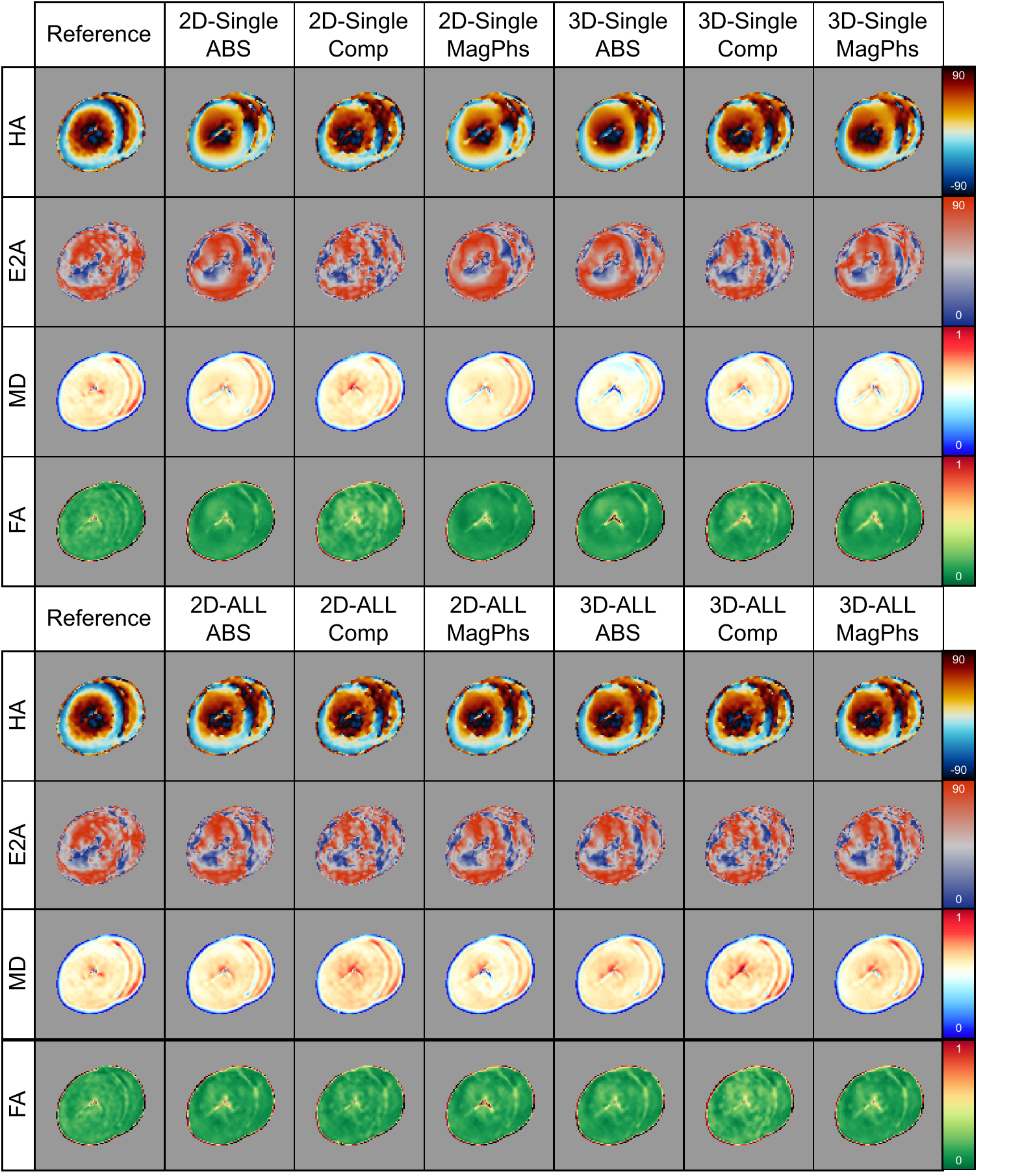}
        \caption{Example DTI maps for an example from the test set.}
        \label{fig:images}
    \end{figure}

    We also visually report an example chosen from the test set in Figure \ref{fig:images}.
    
\section{Discussion}
    Analysing Tables \ref{tab:numerical-results-maps} and \ref{tab:numerical-results-images}, we can see how simpler 2D models that only use magnitude data seem to be extremely stable and easy to train, resulting in very good performance overall. Moreover, these models were also faster to train compared to 3D or complex models. 2D MagPhs models also performed remarkably well given how little the architectures differs from the 2D real-values model. 
    
    Fully complex and 3D models are significantly slower to train on average due to the higher number of operations needed. Moreover, they are also associated with worse performance especially for the DTI maps results.
    
    There is an important point to be made about the worse performances of the more advanced models: as we aimed to keep the number of parameters fixed, we reduced the number of channels for each layer in the more advanced models, reducing the effective capacity of these models and therefore negatively affecting their performances. It can be hypothesised that these more advanced models would perform better given a higher number of fixed parameters and a longer training time. 
    
    Moreover, as our dataset was acquired with SMS factor equal to 2, the 3D architectures have little additional spatial information to exploit when removing the artefacts from the slices. If the acceleration factor were higher, it is likely that better exploiting the 3D spatial information would produce better performances.
    
    Visually, from Figure \ref{fig:images}, we can notice how the AI-derived maps all show extremely similar flaws regardless of the model used (e.g. top-left quadrant of the HA maps), suggesting that 1. the information learned by the models is similar, and 2. almost none of the models is able to overcome the incorrect information present in the SMS version of the images used as input.

\section{Conclusion}
    As cardiac DTI becomes closer and closer to a clinical reality, deep learning is also becoming a vital tool in alleviating some of its downsides such as long scan times and low SNR. In the hurry associated with a new emerging field, many authors prioritise proof-of-concept work to showcase innovative ideas over exploring and comparing known and common options. In this work we lay out the basis of a comparison between input types and model dimensionality and we show how, despite our initial assumption, given a fixed number of parameters and a reasonable train time, 2D models vastly outperform their 3D counterparts and complex-valued networks are not preferable. On the other hand, the seemingly naive use of separate input channels for magnitude and phase data has beneficial performance over discarding the phase information as it is commonly done.
    
    As an advice for future studies, we suggest the use of 2D models and, if available, the of phase information together with the absolute information for their model-development starting point.

\clearpage
\bibliographystyle{splncs04}
\bibliography{PhD,manual}



\end{document}